\documentclass{svjour2}                   
\smartqed  
\usepackage{graphicx}
\usepackage{mathptmx}      
\newcommand{\muas}{$\mu^{\prime \prime}$}

\newcommand{\etal}{et al.}
 \journalname{Experimental Astronomy}
\begin{document}

\title{Formation flying for a Fresnel lens observatory mission}


\author{John Krizmanic$^{1,2}$, Gerry Skinner$^3$,  and Neil Gehrels$^2$}

\institute{J. Krizmanic \at
{NASA Goddard Space Flight Center,} \\
{Greenbelt, Maryland 20771 USA}\\
              Tel.: +001-301-2866817\\
              Fax: +001-301-2861682\\
              \email{jfk@cosmicra.gsfc.nasa.gov}
}

\date{Received: date / Accepted: date}

\maketitle

\begin{center}
{\it 
$^1$Universities Space Research Association \\
$^2$NASA Goddard Space Flight Center,  Greenbelt, Maryland 20771 USA \\
$^3$CESR,  9, avenue du Colonel-Roche
31028 Toulouse, FRANCE
}
\end{center}

\begin{abstract}

The employment of a large area Phase Fresnel Lens (PFL) in a gamma-ray telescope offers the potential to image astrophysical phenomena with micro-arcsecond (\muas) angular resolution \cite{Skinner1}.
In order to assess the feasibility of this concept, two detailed studies have been conducted of formation flying missions in which
a Fresnel lens capable of focussing gamma-rays and the associated detector are carried
on two spacecraft separated by up to 10$^6$ km. These studies were
performed at the NASA Goddard Space Flight Center Integrated Mission
Design Center (IMDC) which developed spacecraft, orbital dynamics, and mission profiles.
The results of the studies indicated that the missions are
challenging but could be accomplished with technologies available currently or in the near term.  The findings of
the original studies have been updated taking account of recent
advances in ion thruster propulsion technology.

\end{abstract}

\section{Introduction}

A unique feature of missions using Fresnel lenses for gamma-ray
astronomy is the large distances over which precise
station-keeping of two spacecraft is required. The concept of a
mission using a Fresnel lens for ultra-high angular resolution
gamma-ray astronomy has been studied by the NASA GSFC Integrated
Mission Design Center (IMDC) \cite{IMDC}. The IMDC is a resource dedicated to developing space mission concepts into advanced mission designs by providing system engineering analysis of all the flight systems and subsystems.   The IMDC process involves the mission scientists working with engineers of all the major engineering disciplines, e.g. mechanical, electrical, propulsion, flight dynamics, communications, mission operations, etc., in a highly interactive environment which naturally allows for inter-discipline communications and trade studies.  For the Fresnel mission studies, an additional formation-flying flight dynamics group also participated. 

Two configurations of the
mission were studied,  a definitive Fresnel mission with a $10^6$ km spacecraft separation and a smaller, pathfinder mission with a $10^5$ km focal length.  Each mission configuration completed a dedicated, one week IMDC study. 
After discussing the requirements and assumptions used as input for the IMDC studies, the mission profiles and key findings are summarized along with a re-analysis of the propulsion requirement to include the implications of recent advances in ion thruster technology.

\section{Mission Requirements and Assumptions}

\begin{table}
  \caption{The parameters for the science payloads for the Full and Pathfinder Fresnel missions.}
  \label{table1}
    \centering
\begin{tabular}{lccl}
\\
\hline \\
        &  Pathfinder   &  Full         &  \\
 \hline \\
Focal length    &     $10^5$  &   $10^6$  &    km\\
Angular Resolution &    $\sim 10$    &       $\sim 1$     &       \muas                \\
Fresnel Lens Diameter                &        1              &          4.5          &    meter \\
Fresnel Lens Mass & 30 & 600 & kg \\
CZT Detector Size                       &   1    &   1  &     meter$^2$    \\
Detector Mass & 200 & 200 & kg \\
Detector Power$^\dagger$ & 200 & 200 & watts \\
CZT Pixel Size & $2\times 2$ & $2 \times 2$ & mm$^2$\\
Pixel Angular Resolution & 4 & 0.4 & \muas \\
Telescope field-of-view & 2000 & 200 & \muas \\
Time on Target & 21 & 21 & days \\
Mean repointing time   &  {\it minimize}    & {\it minimize}    &      \\
 \\ 
\multicolumn{3}{l}{$\dagger$ The power available for detector operation will be much larger due to that} \\
\multicolumn{3}{l}{needed by the ion propulsion for target re-orientation.}
 \\ \hline
\end{tabular}
\end{table}

 The overarching philosophy inherent in both of these IMDC studies was that only technologies already existing or anticipated in the near future were considered as available resources for the spacecraft and mission development.  Thus the results indicate if the missions are feasible in the near term, while identifying areas that need to be improved or developed.

Table 1 details the parameters that were used to define the science payloads.
Starting with the full Fresnel mission, two 4.5 meter diameter aluminum Fresnel lenses in an orthogonal arrangement were incorporated into the lens-craft.  The lenses would be designed for different photon energies and require a simple, $90^\circ$ rotation to exchange one lens for the other.  The size of the Fresnel lenses were reduced to 1 meter in diameter for the pathfinder configuration.  The detector, located on a separate detector-craft,  was chosen to be a 1 meter$^2$ CZT array, segmented into $2 \times 2$ mm$^2$ pixels.  This allowed for the experience with the SWIFT mission \cite{SWIFT} to easily define the detector system parameters, e.g. mass and power.   A single lens-craft and a single detector-craft were initially chosen for each mission configuration along with the desire to accommodate these in a single launch vehicle with an existing defined dual-manifest capability.  

\subsection{Flight Dynamic Parameters}

The choices of orbits and and spacecraft alignment orientations are dictated by the objective of minimizing the propulsion requirements while allowing for the potential viewing of astrophysical objects in any direction, albeit at different times of the orbit cycle.  The on-target time was chosen {\it a priori} to be 3 weeks with the goal of minimizing re-orientation times between observational targets.  A representative list of potential astronomical sources of gamma-rays determined the average re-orientation to be 20$^\circ$.
   
   The spacecraft control and alignment requirements are determined by the performance of the telescope.  The gamma-ray source measurements are insensitive to modest ($\sim 1^\circ$) tilts of either the lens or detector and to mild changes in the separation of the two spacecraft (a 1000 km displacement leads to a $\Delta E = 0.1\%$ for a $10^6$ km baseline).  However, position control and knowledge are essential in the lateral direction, especially in a narrow field-of-view system.  Attitude control must ensure that the image produced by the optics falls onto the detector, and the knowledge of this positioning must be at the level of the size of a detector pixel.  Using a meter size scale for the detector with 2 mm size pixels, the control of the spacecraft needs to be accomplished at the $\sim 10$ cm level with knowledge of $\sim 1$ mm.
   
\section{IMDC Mission Profile}

The challenging aspects of the Fresnel mission are the flight dynamics and satisfying the propulsion requirements.
Flight dynamic analysis led to the selection of an Earth-trailing heliocentric orbit at 1 AU for both missions.  The lens-craft would be in a true orbit while the detector-craft would be in an appropriately offset pseudo-orbit and thus require station-keeping propulsion.  The pseudo-orbit can be within or outside the ecliptic plane depending on the direction of the gamma-ray source. Other orbits, such as those around Lagrange points, were considered but the analysis was determined to be too complex for these studies.
 
 Table 2 details the spacecraft and flight dynamic parameters developed from the IMDC studies.  For the full mission, a Delta-IV-Heavy was identified as the launch vehicle.  The availability of a 19.2 m tall, 5 meter diameter dual-manifest fairing naturally fits the full Fresnel mission profile.  The Delta-IV-H can also throw $\sim 9300$ kg outside the Earth's gravitational potential.  For the pathfinder mission, a smaller Delta-IV-M has sufficient capabilities to achieve the launch although a mission specific dual-payload-attachment-fairing (DPAF) would have to be developed.

\subsection{Propulsion Requirements}

Ion thrusters were identified as the propulsion technology to accomplish both the station-keeping and the re-pointing between observational targets.   The ion thrusters would be incorporated into the detector-craft while both spacecraft would have cold-gas propulsion for fine attitude control.  Solar sail propulsion was considered but dismissed due to the large ($\sim 1$ km$^2$) sail size required and the mission constraints imposed by the inherent directional asymmetry in using the solar flux.

The propulsion requirements were developed from the analysis of the mission flight dynamics.  The results are summarized by detailing the accelerations, $\Delta v$, and thrust requirements, which were evaluated in the context of the available ion thruster performance data, to form each mission profile.  As there is some freedom in the propulsion parameterization, the pathfinder mission will be described using the propulsion performance assumed for the IMDC study while the full Fresnel mission will be considered taking into account recent improvements in ion thruster performance.   The power available for the propulsion is limited to that which can be delivered by the nearly 50 m$^2$ solar arrays assumed in the IMDC design of the detector-craft and corresponds to 19.8 kwatts peak power.

The flight dynamic analysis for the Fresnel pathfinder mission determined that the
accelerations needed for station keeping in an Earth-trailing heliocentric pseudo-orbit varied between $4 - 8 \times 10^{-6}$ m/s$^2$ over the orbit cycle.  Conservatively assuming a mean acceleration of $7 \times 10^{-6}$ m/s$^2$  translates into a $\Delta v_{SK} = 4.2$ m/s/week and an $8.4$ mNewton thrust requirement for a 1200 kg detector-craft. Assuming a two week time to repoint 20$^\circ$ between targets, an acceleration of $10^{-4}$ m/s$^2$ and a $\Delta v_{RP} = 120$ m/s are needed.  The  value of $\Delta v_{RP}$ includes the effects of having to start and stop the detector-craft and using continuous versus impulsive repointing maneuvers. 

The full Fresnel mission requires the spacecraft separation distance to be increased by a factor of 10 to a focal length of $10^6$ km.  Thus the required station-keeping acceleration and $\Delta v_{SK}$ also increases by 10 over that needed for the pathfinder mission.  Assuming a 3 week repoint time and 20$^\circ$ between targets, an acceleration of $4.5 \times 10^{-4}$ m/s$^2$ and a $\Delta v_{RP} = 830$ m/s are required.  An interesting result from the flight dynamics analysis is that the effects of solar gravity do not significantly increase the repointing acceleration requirement for times less than $\sim 3$ weeks.

Table 2 details the flight dynamical requirements and the propulsion parameters for specific configurations of the pathfinder and full Fresnel missions.  For the pathfinder mission, the propulsion parameters for the ion thrusters assumed in the IMDC study are used.  The station-keeping is performed via a single RIT-10 thruster \cite{RIT10} while the repointing maneuvers require the addition of an NSTAR ion thruster \cite{NSTAR}.  Both of these thrusters have flight heritage and have demonstrated $> 20,000$ hours of operation with specific impulses (ISP's) $>3000$ s.  The propulsion power requirements assumed 650 watts for the RIT-10 and 4 kwatts for the NSTAR thruster.  It was noted by the IMDC that this configuration allows for the size of the solar arrays to be reduced to 16 meter$^2$.

For the full Fresnel mission, a re-analysis of the propulsion has been performed, and verified by the IMDC, incorporating recent advances in ion thruster technology.  The
RIT-22 thruster \cite{RIT22} has demonstrated an ISP $> 6000$ s with thrusts up to 250 mN in the laboratory.  From the data in Table 2, the station keeping can be accommodated with a single RIT-22 while the repointing could be accomplished with 4 RIT-22, assuming the thrust could be increased to 300 mN per thruster.  The power requirement assumed 5.5 kwatts per thruster which would require increasing the solar array size by $\sim 10\%$ from the 50 m$^2$ assumed in the original IMDC detector-craft definition.

Although not presented in Table 2, employing RIT-22 thrusters in the pathfinder configuration would allow the repointing to be accomplished on a week timescale.  Thus 60 targets could be viewed in a 5 year mission using 335 kg of fuel with a power requirement of 16.5 kwatts.  A single RIT-10 thruster would still be employed for station-keeping propulsion.

The availability of the improved ion thruster performance allows mass savings such that one could consider including a second detector craft in the full Fresnel mission profile.   The addition of a third spacecraft would allow observations with one detector-craft while the other is maneuvering for the following observation.  Using the data presented in Table 2, the total mass for a single lens-craft and two detector-craft leads to a total of 6150 kg, which is well below the capability of a Delta-IV-H to launch a payload outside the Earth's gravitational potential, even with the customary 20\% contingency.  However, the feasibility of integrating the three spacecraft with a custom payload attachment structure needs to be assessed.

\begin{table}
  \caption{The parameters developed from the IMDC analysis for the Full and Pathfinder Fresnel missions.}
  \label{table1}
    \centering
\begin{tabular}{lccl}
\\
\hline \\
        &  Pathfinder   &  Full         &  \\
 \hline \\
Lens-craft Dry Mass   &     370  &   940  &   kg \\
Lens-craft Power &    260    &       260    &       watts                \\
Detector-craft Dry Mass &            1115              &          1285          &    kg \\
Detector-craft Power & 4.65 & 22 & kwatts \\
Station-keeping $\Delta v$                  &   4.2    &   42  &     m/s/week    \\
On-target time & 21 & 21 & days \\
Station Keeping Thruster ISP & 3000 & 6000 & s \\
Station Keeping Trust$^\dagger$ & 7.8 - 10 & 90 - 185 & mN \\
Mean repointing time & 7 & 21 & days \\
Mean repointing  $\Delta v$ & 120 & 830 & m/s \\
Repointing Thruster ISP & 3000 & 6000 & s \\
Repointing Thrust$^\dagger$ & 110 - 145 & 580 - 1200 & mN \\
Total Propellant Mass & 320 & 1320 & kg \\
\# of targets in 5 years$^\ddagger$ & 52 & 43 & \\
\\
\multicolumn{3}{l}{$\dagger$ The thrust range is defined by the end-of-life (dry) mass versus the initial} \\
\multicolumn{3}{l}{(wet) spacecraft mass.} \\
\\
\multicolumn{3}{l}{$\ddagger$ If the number of detector-craft is increased to 2, the number of targeted } \\
\multicolumn{3}{l}{ observations could be increased up to a factor of 2.}

 \\ \hline
\end{tabular}
\end{table}

\subsection{Attitude control}

The orientation of the lens spacecraft is not critical (60"
control and 10" knowledge were assumed, though even though these
requirements could be further relaxed). Commercially available
laser gyro, star-tracker and reaction wheel units having the
necessary capabilities were identified in the IMDC studies.

The intrinsic requirements for the detector- and lens-craft alignment in the two Fresnel configurations are
similar. 
The formation flying alignment could be accomplished by employing
a laser beacon on the lens-craft, which then would be observed by a sensor
on the
detector-craft. This would require a sensor working at the micro-
arcsecond level for the full Fresnel mission, with perhaps a 100 microarcsecond field of view.
One of the options considered was to place the sensor on a gimbal
to allow fine pointing without placing undue constraints on the
spacecraft orientation.

A key finding of the IMDC study was the need for further development of a metrology system to determine the absolute position of the gamma-ray sources in the narrow field-of-view of the telescope.  The employment of a micro-arcsecond star tracker was determined to be problematic as it required meter-sized apertures and excessive integration times or a star of sufficiently large magnitude in the field-of-view.  A gimbaled micro-arcsecond star tracker of more modest size viewing bright, off-axis stars operating with precise gyros was identified as a possible solution. However, this technology needs to be further developed.  Other missions, such as MAXIM, have similar metrology requirements and could offer a solution to this problem \cite{MAXIM}.  Another intriguing possibility considered by the IMDC was to include the deployment of another detector-craft at a much shorter separation, from the lens-craft, to act as a finder-craft.  The widened field-of-view of this telescopic configuration relaxes the source-finding problem.  The alignment of the three spacecraft places the astrophysical source in the field-of-view of the longer, baseline configuration, once the finder-craft drifts out of the field-of-view.  However, this bootstrapping technique requires the addition of a third, maneuverable spacecraft, albeit with modest propulsion requirements.

\subsection{Spacecraft, thermal control, telemetry}

Other aspects of the IMDC study, such as spacecraft mechanical and electrical design, thermal control and management, telemetry, etc., were found by the IMDC studies to be 
straightforward.  Communications could be handles 
as an S-band link between the two spacecraft for
ranging, with the data from the lens-craft being
relayed through the detector-craft.  A 1.5 m gimbaled
antenna then would allow all data from both spacecraft to be
sent to the ground during a single daily 15 minute DSN contact.  It was noted that halting the Earth-drift-away orbit at $\sim 0.1$ AU would preclude any obscuring of the spacecraft by the sun.

\section{Conclusions}

Two gamma-ray astronomy missions employing Fresnel lenses were developed at the NASA GSFC Integrated Mission Design Center; a pathfinder mission with a $10^5$ km spacecraft separation and 10 micro-arcsecond (\muas)  imaging ability and a definitive mission with a
$10^6$ km separation and with a 1 \muas~  angular resolution.  The development of the spacecraft was determined to be straightforward. While the studies determined that the flight dynamics and propulsion requirements are challenging, they can be accomplished with current ion propulsion technology if a re-pointing time scale of a week is used.  Furthermore, recent advances in ion thrusters have relaxed the propulsion requirements potentially allowing the incorporation of an additional detector-craft into a mission profile, thus doubling the number of viewed gamma-ray sources in a 5 year mission lifetime.  This work was supported by a grant from the NASA Revolutionary Aerospace Systems Concepts (RASC) program.



\end{document}